\documentclass[amsmath,amssymb, aps, pra, twocolumn,superscriptaddress]{revtex4-2}

\usepackage{float}
\usepackage{graphicx}
\usepackage{dcolumn}
\usepackage{bm}
\usepackage{hyperref}
\hypersetup{
    colorlinks=true,
    linkcolor=blue,
    filecolor=magenta,      
    urlcolor=cyan,
}
\usepackage{color}
\usepackage{fancyhdr}
\usepackage[normalem]{ulem}
\usepackage{physics}
\usepackage[dvipsnames]{xcolor}

\definecolor{greenishred}{rgb}{0.6, 0.4, 0}

\begin{document}

\title{Realizing the Frenkel-Kontorova model with Rydberg-dressed atoms}

\author{Jorge Mellado Mu\~noz}
\affiliation{School of Physics and Astronomy, University of Birmingham, Edgbaston, Birmingham, B15 2TT, UK}
\author{Rahul Sawant}
\email{r.v.sawant@bham.ac.uk}
\affiliation{School of Physics and Astronomy, University of Birmingham, Edgbaston, Birmingham, B15 2TT, UK}
\author{Anna Maffei}
\affiliation{School of Physics and Astronomy, University of Birmingham, Edgbaston, Birmingham, B15 2TT, UK}
\affiliation{Dipartimento di Fisica “E. Fermi”, Universit\`a di Pisa, Largo B. Pontecorvo 3, 56127 Pisa, Italy}
\author{Xi Wang}
\affiliation{School of Physics and Astronomy, University of Birmingham, Edgbaston, Birmingham, B15 2TT, UK}
\author{Giovanni Barontini}
\affiliation{School of Physics and Astronomy, University of Birmingham, Edgbaston, Birmingham, B15 2TT, UK}

\date{\today}
\begin{abstract}
We propose a method to realize the Frenkel-Kontorova model using an array of Rydberg dressed atoms. Our platform can be used to study this model with a range of realistic interatomic potentials. In particular, we concentrate on two types of interaction potentials: a springlike potential and a repulsive long-range repulsive. We numerically calculate the phase diagram for such systems and characterize the Aubry-like and commensurate-incommensurate phase transitions. Experimental realizations of this system that are possible to achieve using current technology are discussed. 
\end{abstract}
\maketitle
\section{Introduction}
The Frenkel-Kontorova model (FKM) was introduced to describe the structure and dynamics of a crystal lattice near a dislocation core. It consists of a chain of particles with long-range spring interactions placed in a sinusoidal potential. Its main characteristic is the competition between the two length scales promoted by these two potentials. Depending on whether the ratio between the interparticle distance and the substrate period is rational or irrational, the particle configuration becomes \emph{commensurate} or \emph{incommensurate} with the substrate. A system in the incommensurate phase can undergo the so-called Aubry phase transition, characterized by the transition from an \emph{unpinned} to a \emph{pinned} configuration when the strength of the substrate potential is increased~\cite{braun2013frenkel}. This transition is identified by a change in the particle positions and in the phonon spectrum of the ground state configurations. The FKM has proven useful to describe a multitude of condensed matter systems. Some examples are the study of dislocation dynamics in solids~\cite{fitzgerald1991dislocations}, surfaces and adsorbed atomic layers~\cite{braun1989interaction}, incommensurate phases in dielectrics~\cite{blinc1986incommensurate}, crowdions~\cite{xiao2003adatom}, magnetic chains~\cite{cowley1976neutron}, Josephson junctions~\cite{mclaughlin1978perturbation} and tribology~\cite{vanossi2013colloquium} using the Frenkel-Kontorova-Tomlinson model ~\cite{weiss1996dry,weiss1997dry}. 

\begin{figure}[t]
    \centering
    \includegraphics[width=0.4\textwidth]{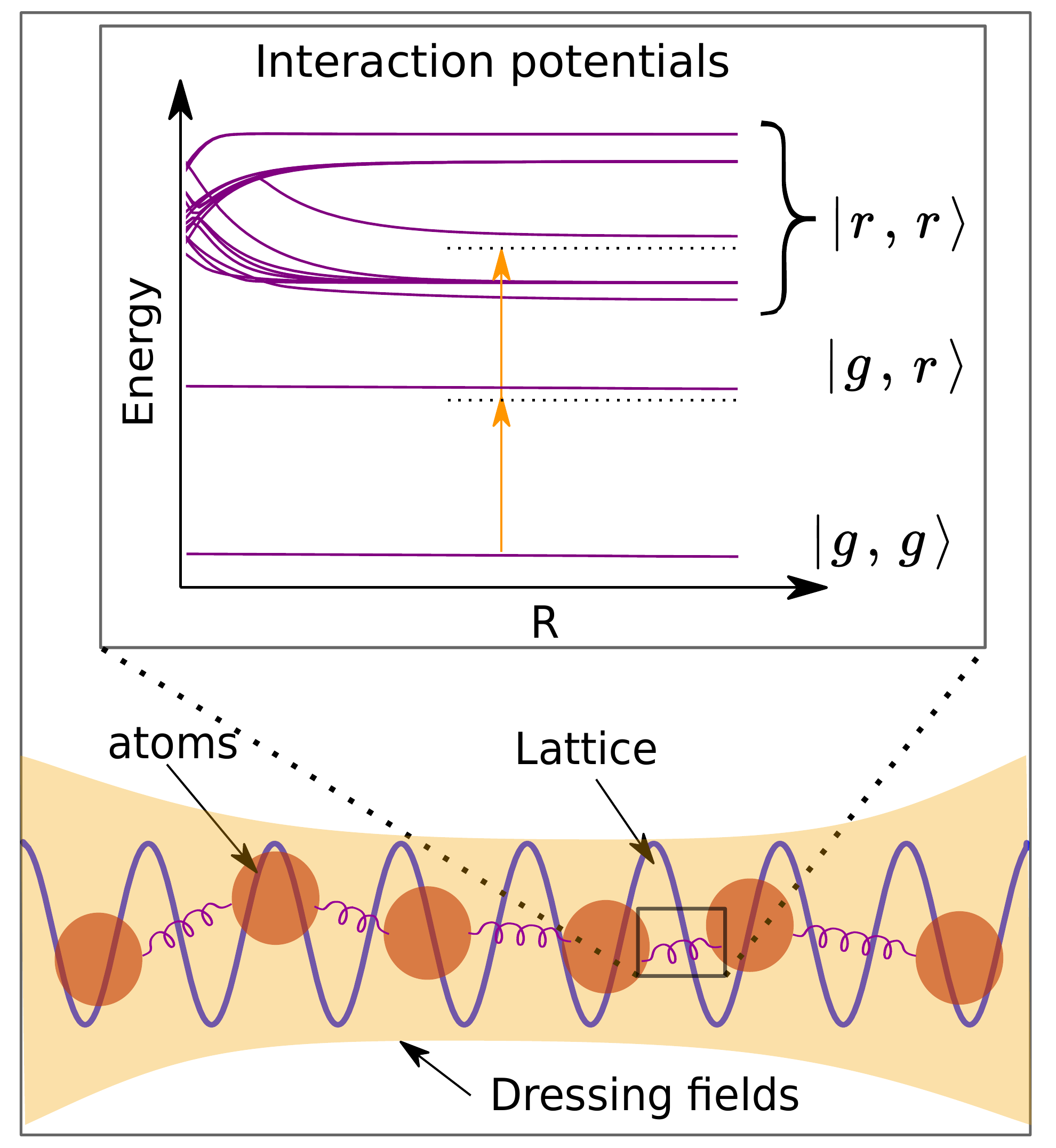}
    \caption{Schematic of the system for the realization of the Frenkel-Kontorova model. An ensemble of atoms in a 1-D optical lattice interacts via Rydberg-dressed potentials. The inset shows how the interatomic potentials can be realized by dressing the atoms' ground state with the Rydberg levels.   }
    \label{system_figure}
\end{figure}

Although effective, the FKM uses a non-realistic infinite-range spring interaction potential. It is therefore beneficial to develop a fully controlled system where the effect of realistic interaction potentials can be tested. Cold atoms systems are ideal candidates for this purpose due to their high degree of control and flexibility. The introduction of optical lattices allowed for the study of several models that explain different condensed matter phenomena such as the Superfluid-Mott insulator phase transition~\cite{jaksch1998cold,greiner2002quantum}, Anderson localization~\cite{roati2008anderson,billy2008direct} and the effects of quantum magnetism~\cite{Greif_Short2013}. Furthermore, long-range interactions can be achieved using atomic species with high permanent dipolar moments~\cite{lu2011strongly}, dipolar molecules~\cite{blackmore2018ultracold}, ultracold ions~\cite{bylinskii2015tuning} or Rydberg-dressed atoms~\cite{pupillo2010strongly}. In particular, Rydberg-dressed atoms have the advantage of enabling the control of the range and functional dependence of the interaction potentials, using various Rydberg states for the dressing. These dressed states are achieved by weakly admixing excited Rydberg states with the ground state, using near-resonant light~\cite{zeiher2016many} (see Fig.~\ref{system_figure}).

In this work, we propose the implementation of the FKM with cold Rydberg dressed atoms in an optical lattice. We show that, by using Rydberg dressing and realistic experimental parameters, it is possible to realize at least two different variants of the FKM. As shown in section~\ref{sec:system}, we concentrate in particular on a springlike interaction potential, similar to the original FKM, and a repulsive potential. In section~\ref{sec:diagram} we calculate the phase diagrams of the system for both cases, which feature the characteristic incommensurate and commensurate configurations. We show how the equilibrium configurations take the form of different \emph{devil's staircases} as the amplitude of the lattice potential is varied. In section~\ref{sec:aubry} we concentrate on the system in the incommensurate configuration. We show that, depending of the interaction potential, it is possible to observe either a Aubry-like transition or a crossover from an unpinned phase to a pinned phase. Interestingly, the crossover is characterized by the excitation of a \emph{soft mode}, resembling the phason mode typical of infinite systems. In Section V we report our conclusions. 

\section{The system}
\label{sec:system}
Our system, which is depicted in Fig.~\ref{system_figure}, consists of $N$ atoms arranged in a 1D chain placed in a tunable optical lattice, and a dressing field that produces the Rydberg dressed states. We limit the interactions to nearest neighbors since the densities are such that any n-body potential is a sum of 2-body potentials~\footnote{We checked this explicitly in our case.}. Indicating with $x_i$ and $P_i$ the particle positions and their momenta, the energy of the system is
\begin{equation}
\begin{split}
    E & = \sum\limits_{i=1}^{N} \frac{P^2_i}{2m} + \frac{1}{2} \sum\limits_{i=1}^{N-1} c_1V_{\text{int}}(x_{i+1}-x_{i}) \\ 
    & +  \sum\limits_{i=1}^{N}  \left[V_0-V_0\cos \left( 2 \pi x_i/b \right) \right],
    \label{Eqn:energy_eqn}
\end{split}
\end{equation}
where $V_{\text{int}}(x_{i+1}-x_{i})$ is the normalized nearest neighbour interaction between the particles, $c_1$ is a coefficient that accounts for the amount of Rydberg admixture in the dressed state, $V_0$ is the depth of the lattice potential and $b$ is the lattice spacing. $m$ is the mass of the atoms.

\begin{figure}[t]
    \centering
    \includegraphics[width=0.45\textwidth]{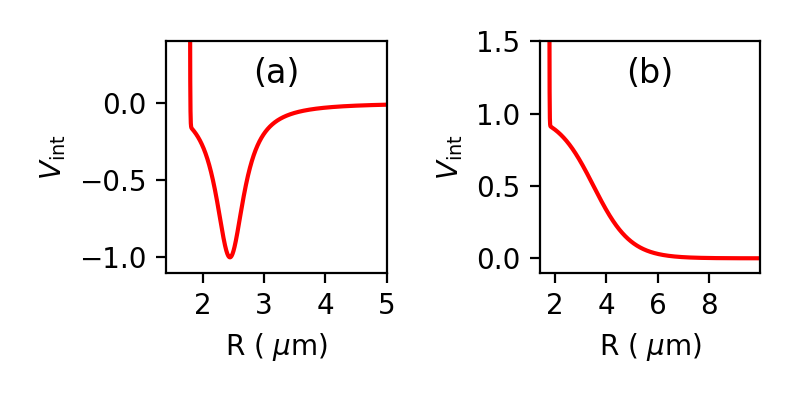}
    \caption{Normalized interatomic potentials obtained by dressing the atoms with Rydberg states, as explained in the text. (a) The springlike potential  (b) The repulsive potential. }
    \label{fig_ryd:fits_only}
\end{figure}

We study the system for two different interaction potentials that can be realistically implemented. One is a long-range \emph{springlike} potential and the other a \emph{repulsive} potential, as depicted in Fig.~\ref{fig_ryd:fits_only}~(a) and \ref{fig_ryd:fits_only}~(b) respectively. The derivation of these potentials from the Rydberg spectrum is shown in appendix~\ref{ryd_dress_appendix}. The functional form of the \emph{springlike} potential is 
\begin{equation}
    V_\mathrm{int} = \frac{1}{\left(1+e^{c_0 (r-r_0)}\right)}+ \frac{e^{-c_2 (r-r'_{0})}}{c_1},
\end{equation}
where $r_0$, $c_0$, $c_2$ and $r'$ are parameters that depends on the details of the Rydberg dressing. The shape of the \emph{repulsive} potential is instead given by:
\begin{equation}
 V_\mathrm{int} =\frac{-1}{\left(1+c_0 (r-r_0)^2\right)}+ \frac{e^{-c_2 (r-r'_{0})}}{c_1}.
\end{equation}

In order to give a specific example, we choose $^{87}$Rb atoms dressed with the Rydberg levels $60S_{1/2}$ or $60P_{1/2}$, to realize the two interaction potentials, repulsive or springlike respectively (details are in the appendix~\ref{ryd_dress_appendix}). The dressing of the atoms can be realized by a two-photon transition in the case of $60P_{1/2}$, using two lasers at 794.98 nm and 480.21 nm for the $5S_{1/2} \rightarrow 5P_{3/2}$ and $5P_{3/2} \rightarrow 60S_{1/2}$ transitions respectively. While the dressing using $60P_{1/2}$ can be realized by a single photon transition, $5S_{1/2} \rightarrow 60P_{1/2}$, addressed by a 297.11nm laser. With these parameters we have $r_0 = 3.494$~$\mu$m, $c_0 = 1.382$~$\mu$m$^{-1}$, $c_1/k_B = 248.9$~nK, $c_2 = 145$~$\mu$m$^{-1}$ and $r'_{0} = 1.811$~$\mu$m for the springlike potential and $r_0 = 2.440$~$\mu$m, $c_0 = 12.951$~$\mu$m$^{-1}$, $c_1/k_B = 783.6$~nK, $c_2 = 263$~$\mu$m$^{-1}$, $r'_{0} = 1.8127$~$\mu$m for the repulsive potential.

Concerning the tunable optical lattice, it is possible to realize it by interfering two light beams at an angle. This angle can then be varied to change the lattice spacing~\cite{Li_atomic_2008,Ville_loading_2017} allowing for lattice periods in the desired range, which in this work we choose to be $1.9-4.5~\mu$m. The system that we propose can be practically implemented. Indeed Rydberg-dressed atoms have proven to be stable against losses in an optical lattice~\cite{Macri_rydberg_2014}. 

\section{The phase diagrams}
\label{sec:diagram}
\begin{figure*}[t]
    \centering
    \includegraphics[width=0.95\textwidth]{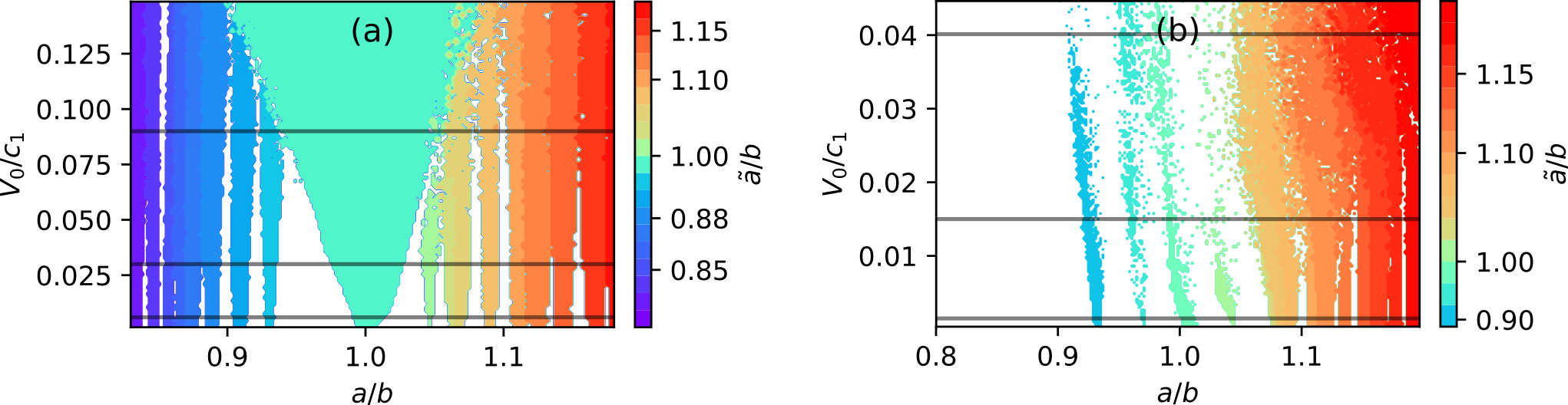}
    \caption{Phase diagram showing $\tilde{a}/b$ as a function of $a/b$ and $V_0/c_1$ for (a) the springlike potential and (b) the repulsive potential. Each color represents a different lock-in region, characterized by a commensurate configuration where $\tilde{a}/b$ is rational.}
    \label{FK_Fig:phase_diagram}
\end{figure*}
In this section, we compute the phase diagram for the ground state of the system for both the dressed potentials. As mentioned above, the FKM is characterized by two length scales, which in our case are the lattice spacing $b$ and the distance $a$ that minimizes $V_{int}$. The mean interatomic distance $\tilde{a}$ that minimizes $E$ when $V_0\neq0$ will, therefore, result from the competition between these two length scales. In particular, depending on which value $\tilde{a}$ takes, the phase diagram breaks between commensurate and incommensurate phases. In a commensurate configuration, the positions of the atoms can be expressed as $x_{Q+i} = x_i + R\cdot b$, where $Q$ and $R$ are integers and $i$ denotes the index of the particle. Therefore the mean interparticle distance is the rational number $\tilde{a}/b = R/Q$. An incommensurate configuration is instead characterized an irrational value of $\tilde{a}/b$. 

To compute the ground state and derive the phase diagrams, we use the generalized simulating annealing algorithm \cite{Tsallis_generalized_1996}. We find the minimum of the energy functional (\ref{Eqn:energy_eqn}) for different values of $V_0$ and $b$, with the condition that all the particles are at rest ($P_i=0$). For both cases, we consider a system of $N=50$ atoms. In the repulsive case, we add hard walls to confine the system and prevent the particles from separating indefinitely. Such hard walls could be realized using the technique of reference ~\cite{gaunt2013bose}. We chose the distance between the confining walls so that $a=3.17~\mu m$. In the springlike case, $a$ is instead fixed by the minimum of the interaction potential, which in our specific case is at $a=2.44~\mu m$. 

The resulting phase diagrams are shown in  Fig.~\ref{FK_Fig:phase_diagram}, where we report $\tilde{a}/b$ as a function of $a/b$ and $V_0$. In particular, the coloured regions indicate the largest \emph{lock-in regions} for $\tilde{a}/b$, corresponding to commensurate configurations. The white regions are characterized by smaller lock-in regions and incommensurate configurations. In both phase diagrams, for $V_0=0$ the system is in the so-called \emph{floating phase} where $a/b$ can take any value. As $V_0$ is increased, commensurate configurations become more energetically favorable and the phase diagram starts to break into lock-in regions. Indeed, around each rational value of $a/b$ there are intervals in which $\tilde{a}/b$ take the same rational value. The amplitude of the lock-in regions increases as $V_0$ increases. For the trivial case $V_0/c_1\gg1$ all the particles are pinned in the minima of the lattice potential and therefore only commensurate configurations are possible.
\begin{figure*}[t]
    \centering
    \includegraphics[width=0.9\textwidth]{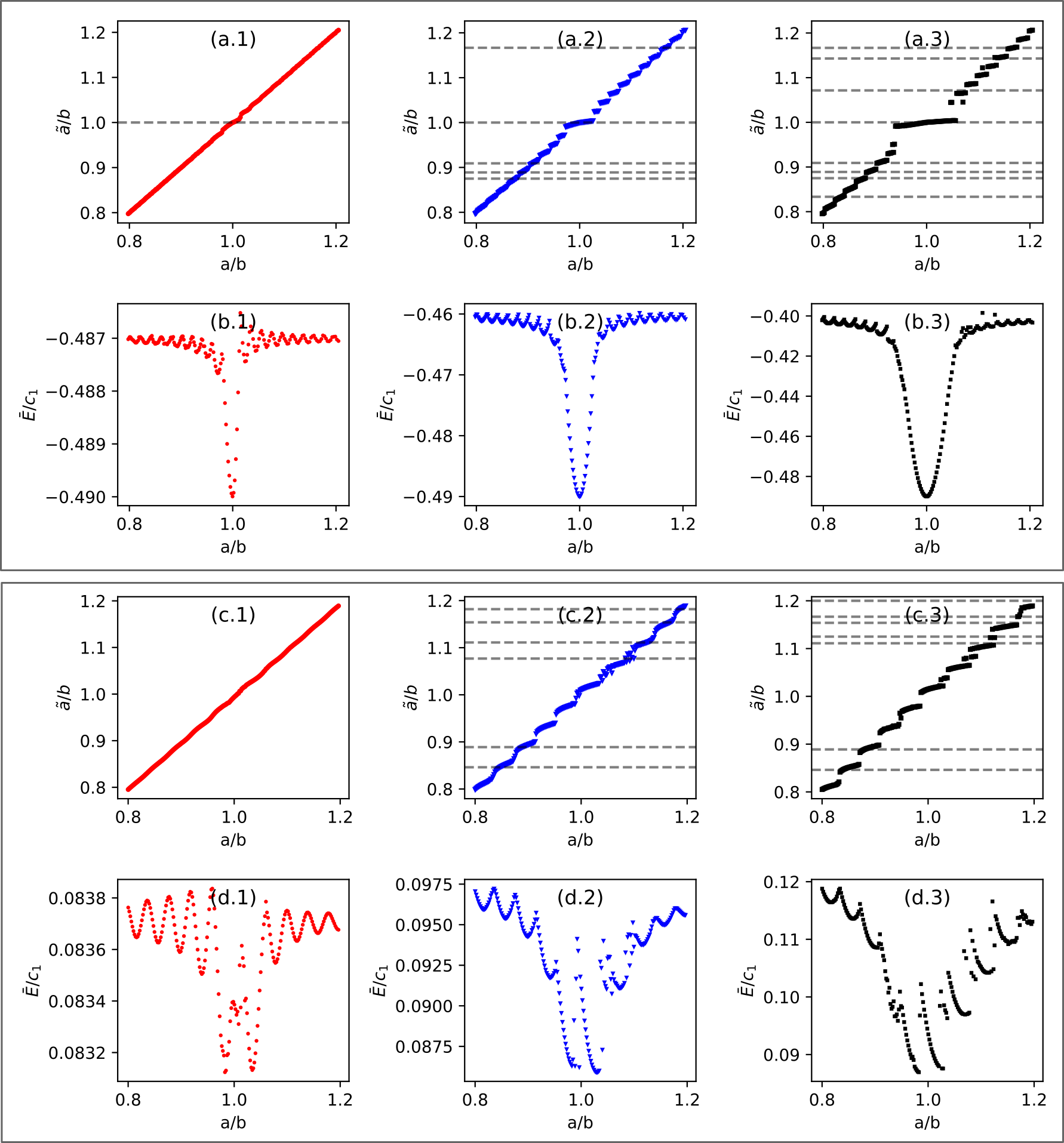}
    \caption{(a.1,2,3) and (c.1,2,3) show the mean interparticle distance with respect to the ratio $a/b$ for springlike and repulsive potentials respectively. (b.1,2,3) and (d.1,2,3) show the energy per particle ($\bar{E}$) with respect to the ratio $a/b$. Here $V_0/c_1 = 0.006, 0.03, 0.09$ and $0.0015, 0.015, 0.04$ for springlike and repulsive potentials respectively. The horizontal dashed lines show the locations of a few rational values of $\tilde{a}/b$.}
    \label{FK_Fig:staicase}
\end{figure*}

Let us first analyze the springlike interaction case, which is the one more similar to the original FKM. In Fig.~\ref{FK_Fig:staicase}~(a) we report $\tilde{a}/b$ as a function of $a/b$ for three values of $V_0/c_1$, indicated as black horizontal lines in Fig.~\ref{FK_Fig:phase_diagram}. 
As can be seen in Fig.~\ref{FK_Fig:staicase}~(a.1), for small but finite values of $V_0/c_1$ small intervals of zero slope start to appear in the curve $\tilde{a}/b$ vs. $a/b$. 
Such intervals are centered around commensurate values of $a/b$ and correspond to $\tilde{a}/b$ taking the same rational value (horizontal dashed lines). As the curve is a combination of zero and non-zero slope regions, it is referred to as an \emph{incomplete devil's staircase}. The lock-in regions exist because the transition from a rational to an irrational value involves the creation of a \emph{discommensuration}, which costs energy. This can be seen in Fig.~\ref{FK_Fig:staicase}~(a.2), where we report the energy per particle as a function of $a/b$. As $a/b$ moves away from a rational value, the energy per particle starts to increase. When this energy exceeds the energy required to create a discommensuration, a transition to an incommensurate phase occurs. As reported in Fig.~\ref{FK_Fig:staicase}~(a.3 and a.4), when $V_0/c_1$ is further increased, the width of the lock-in regions increases and incommensurate configurations start to disappear. Eventually, as shown in Fig.~\ref{FK_Fig:staicase}~(a.4), the curve $\tilde{a}/b$ vs. $a/b$ becomes a \emph{complete devil's staircase}. The transition from an incomplete to a complete devil's staircase is of special interest for some condensed matter systems like, e.g., polymers. Also, in this case, moving away from the center of the lock-in region leads to an increase of the energy per particle, until `jumping' to the next lock-in region becomes energetically favorable.  

The same analysis for the repulsive case is reported in Fig.~\ref{FK_Fig:staicase}~(c) and~\ref{FK_Fig:staicase}~(d). In this case, an \emph{anomalous} incomplete devil's staircase is formed in the function $\tilde{a}/b$ vs. $a/b$. The anomaly is in the fact that the lock-in regions are not characterized by a zero slope, and therefore are a mixture of commensurate and incommensurate configurations. This is reflected also in Fig.~\ref{FK_Fig:phase_diagram}~(b), where a large part of the phase diagram is not covered by commensurate configurations. The anomalous lock-in regions increase as $V_0$ is increased, but the slope remains finite even for large values of $V_0$, therefore the devil's staircase remains always incomplete. Similar to the springlike interaction potential, the lock-in regions are characterized by the increase of the energy per particle as the system moves away from the center of the region.

\section{Incommensurate Configurations}
\label{sec:aubry}
The competition between the lattice potential and the interatomic potential becomes apparent when the two systems are incommensurate with each other, i.e., when the ratio $a/b$ is highly irrational and quite far from a rational value. In this section, we analyze the behavior of such a system. We first look at how the ground state configuration of the particle undergoes a phase transition from an unpinned phase to a pinned phase. The pinned phase is characterized by the particles taking only a handful of specific locations with respect to the lattice potential. We then study the phonon spectrum of the ground state and see how the transition changes with the two dressed interaction potentials. 
In the original FKM, an incommensurate system containing infinite particles can undergo a phase transition called the Aubry phase transition when $V_0/c1$ crosses a critical point~\cite{aubry1983twist}. This can also be interpreted as a transition from an unpinned phase to a pinned one.  Additionally, this type of transition is characterized by the appearance of a gap for the minimum frequency in the phonon spectrum~\cite{floria1996dissipative}. For our finite system of particles, we observe an analogous Aubry-like transition for the springlike case, while we observe a smooth crossover for the repulsive case.

\begin{figure}
    \centering
    \includegraphics[width=0.35\textwidth]{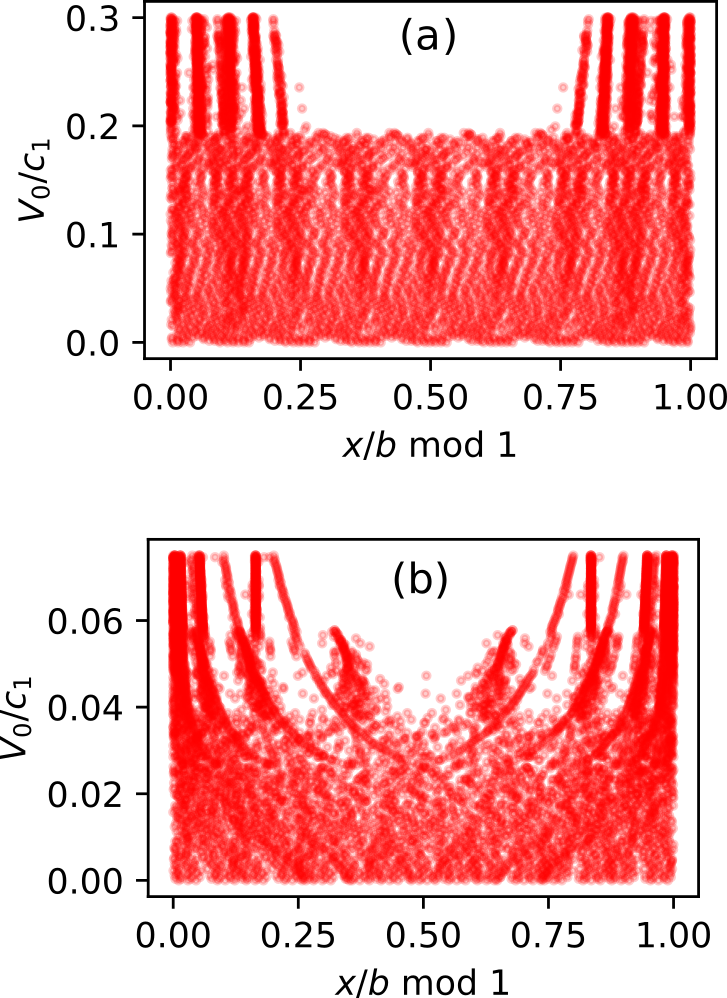}
    \caption{Particle position with respect to the lattice potential for each value of $V_0/c_1$ for (a) the springlike and (b) the repulsive case respectively.  
    Here the system is incommensurate with the ratio $a/b = 0.873$.}
    \label{FK_Fig:aubry}
\end{figure}

In Fig.~\ref{FK_Fig:aubry} we report ${x_i}/{b} \ \text{mod} \ 1$, which gives the particle position with respect to the lattice phase, as a function of $V_0/c_1$, for both the dressed potentials. To provide a specific example, we chose the configuration with $a/b=0.873$. In the trivial case of $V_0/c_1 \approx 0$, the particles are not restricted to any particular position with respect to the lattice. In the springlike case, as $V_0$ is increased, there is a rather abrupt transition to a configuration where the number of allowed positions is drastically reduced. The system is in the unpinned phase until $V_0/c_1 = 0.19$, after which it undergoes a steep transition to the pinned phase. For the repulsive case, we observe a relatively smoother crossover between $V_0/c_1 = 0.03$ and 0.04 from the unpinned phase and the pinned phase, as can be seen in Fig.~\ref{FK_Fig:aubry}~(b).

\begin{figure}
    \centering
    \includegraphics[width=0.5\textwidth]{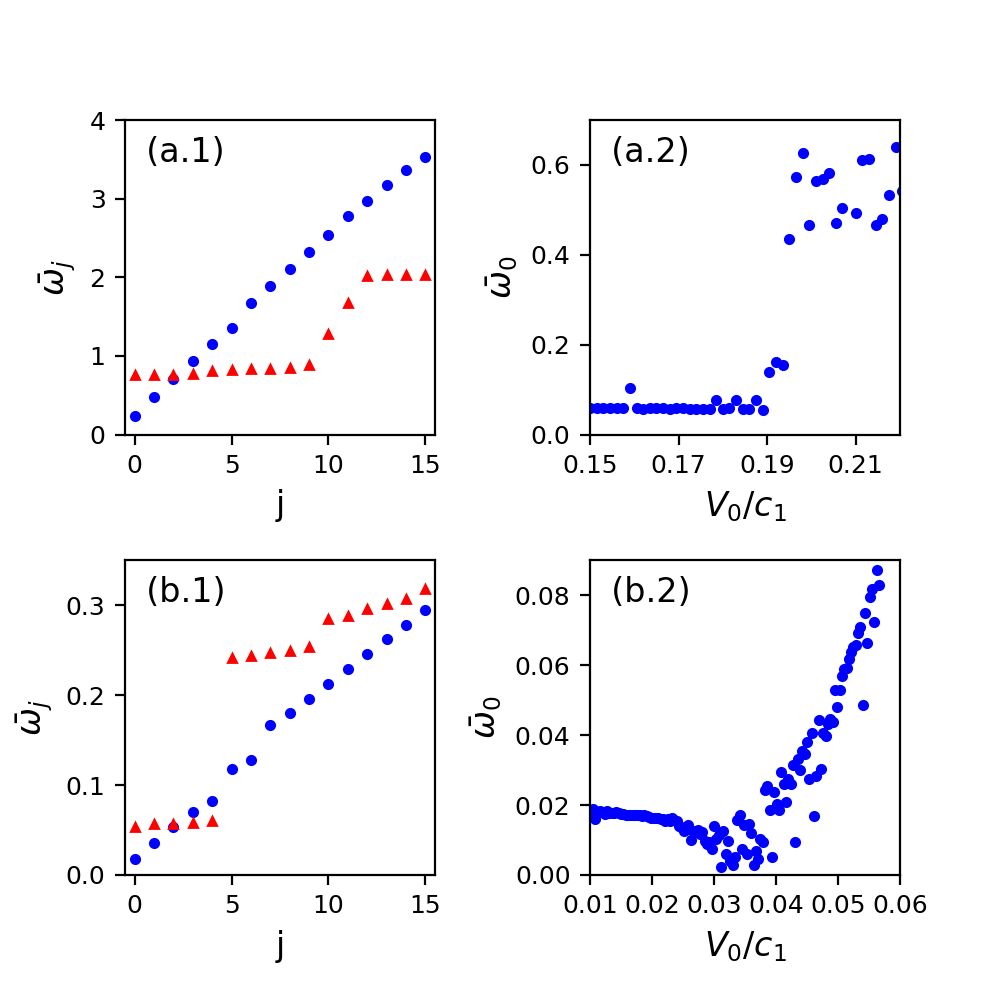}
    \caption{(a.1) Adimensional frequencies for the 15 lowest modes of the phonon spectrum for the springlike case for $V_0/c_1=0.180$ (blue dots) and $V_0/c_1=0.201$ (red triangles). (b.1) is the same for the repulsive case for $V_0/c_1=0.01275$ (blue dots) and $V_0/c_1=0.024$ (red triangles). (a.2) and (b.2) show the frequency of the lowest mode as a function of $V0/c_1$.   
    In all cases the system is incommensurate with the ratio $a/b = 0.873$.}
    \label{FK_Fig:phonon}
\end{figure}

Another signature of the transition can be found in the phonon spectrum of the ground state configurations. An infinite incommensurate system in the unpinned phase presents a gapless Goldston mode called \emph{phason}. The phason is associated to the invariance of the uniform relative translation of the phases of modes with relatively irrational periodicities. As the system undergoes the Aubry transition to the pinned phase, this mode disappears. This is a consequence of the pinning of the particles to the substrate, not allowing for the dynamics of this zero-frequency mode. Since our system is finite, such phason mode does not appear in the unpinned phase. However, as we show here below, for the springlike case we observe a similar kind of transition while, for the repulsive case, a \emph{soft mode} appears in correspondence of the crossover from the unpinned to pinned phase. A soft mode is an excitation above the ground state whose energy vanishes in the limit $N\rightarrow \infty$~\cite{sharma1984aubry}, where it becomes a phason.

We calculate the phonon spectrum of different configurations using the previously calculated ground states and the introduction of the dynamical matrix, defined as,
\begin{equation}
    D_{kl}=\frac{\partial^2 E}{\partial x_k \partial x_l} \Bigr|_{\{ x^G_i \}} ,
\end{equation}
where $D_{k,l}$ are the elements of the dynamical matrix and $\{ x^G_i \}$ is the ground state particle configuration. The eigenvalues of this matrix are $\lambda_j$ for $j=1,N$ and they are related to the frequencies of the possible phonon modes, $\omega_j$, as $\omega^2_j=\lambda_j$. For convenience, we introduce the adimensional phonon frequency, $\bar{\omega}_j$, defined as $\bar{\omega}_j=\omega_j \sqrt{m a^2 / c_1}$ where $m$ is the mass of the particles.

In Figs.~\ref{FK_Fig:phonon}~(a.1) and~\ref{FK_Fig:phonon}~(b.1) we report the frequency of the lowest frequencies modes for the springlike and repulsive cases respectively. The blue dots are for the unpinned phase and the red triangles are for the pinned phase. To provide a specific example, we chose $a/b=0.873$. For both dressed potentials, the minimum frequency is almost zero below the transition, while the gap becomes larger in the pinned phase. In both cases, we can observe the formation of a staircase in the phonon spectrum as $V_0$ is increased. This behavior is similar to the one of the original FKM, in which a staircase formation is also observed in the phonon spectrum in the incommensurate case~\cite{floria1996dissipative}. 

In Figs.~\ref{FK_Fig:phonon}~(a.2) and~\ref{FK_Fig:phonon}~(b.2) we report the lowest energy mode $\bar{\omega}_0$ as a function of $V_0/c_1$, for the springlike and repulsive case respectively. For the springlike case the gap opens up suddenly for $V_0/c_1\simeq0.19$, indicating indeed that the system undergoes the Aubry-like transition from the unpinned to the pinned phase, in agreement with Fig.~\ref{FK_Fig:aubry}~(a). For the repulsive case, instead, the energy of $\bar{\omega}_0$ first decreases until it approaches zero for $V_0/c_1\simeq0.032$. For this value, the system supports a soft mode, similar to the infinite FKM. As $V_0/c_1$ is further increased, the energy of $\bar{\omega}_0$ increases again, and the system crosses-over to the pinned phase. As $a/b$ moves away from an irrational value and approaches a rational one, both the Aubry-like transition for the springlike case and the crossover for the repulsive case happen for lower values of $V_0/c_1$, until a commensurate configuration is reached. 

\section{Conclusions}

In conclusion, we have proposed a system of Rydberg dressed atoms in an optical lattice as a platform for the study of the FKM with realistic potentials. We have reported the phase diagrams for two dressed potentials that can be realized experimentally. We have shown that, depending on the shape of the interaction potential, the system can exhibit different behaviors. In particular, with a springlike potential, the phenomenology is very close to the original FKM, while for a repulsive potential, the system exhibits some anomalous features. This is particularly apparent in incommensurate configurations, where the springlike case undergoes an Aubry-like transition as the height of the lattice is increased, while the repulsive case is characterized by a relatively smooth crossover from the unpinned to the pinned phase.  

The required Rydberg-dressing can be implemented using the details from appendix \ref{ryd_dress_appendix}. In order to find the ground states, the experiment can be performed similarly to the simulated annealing. The atoms can be first loaded into a tight optical lattice and the Rydberg interaction can be switched on. Then the temperature of the atoms can be lowered, either by evaporative cooling or by a bath of colder atoms of a different species co-trapped with the Rydberg-dressed atoms. While the cooling process is going on, the lattice depth can be simultaneously lowered to obtain the value we desire to study. Depending on the amount of Rydberg admixture and therefore the strength of the interactions, the phase diagram can be explored with samples at temperatures that can be achieved in state-of-the-art experiments. 

The proposed system can, therefore, be realized with current technology and can be extended to other kinds of interaction potentials, including attractive ones, using different dressing schemes. It is relatively straightforward to set up moving optical lattices to implement the Frenkel-Kontorova-Tomlinson model and perform tribology studies with unprecedented control. This is promising since the study of this system with ultracold ions, which are limited to ion-ion interactions, has proven both, theoretically~\cite{benassi2011nanofriction,pruttivarasin2011trapped} and experimentally ~\cite{bylinskii2015tuning,bylinskii2016observation,kiethe2017probing,gangloff2020kinks} to be an excellent platform to study nanofriction and other phenomena related to the properties of the FKM. Another extremely interesting direction could be to extend the system in two dimensions, where numerical calculations are difficult, and where an experimental implementation could provide new insight. 

\paragraph*{Acknowledgements}
The authors are supported by the Leverhulme Trust Research Project Grant UltraQuTe (grant number RGP-
2018-266). AM is supported by the Erasmus programme.

\paragraph*{Author Contribution}
JMM and RS contributed equally to this work.


\appendix
\section{Rydberg Dressing}\label{ryd_dress_appendix}
\begin{figure*}[t]
    \centering
    \includegraphics[width=0.75\textwidth]{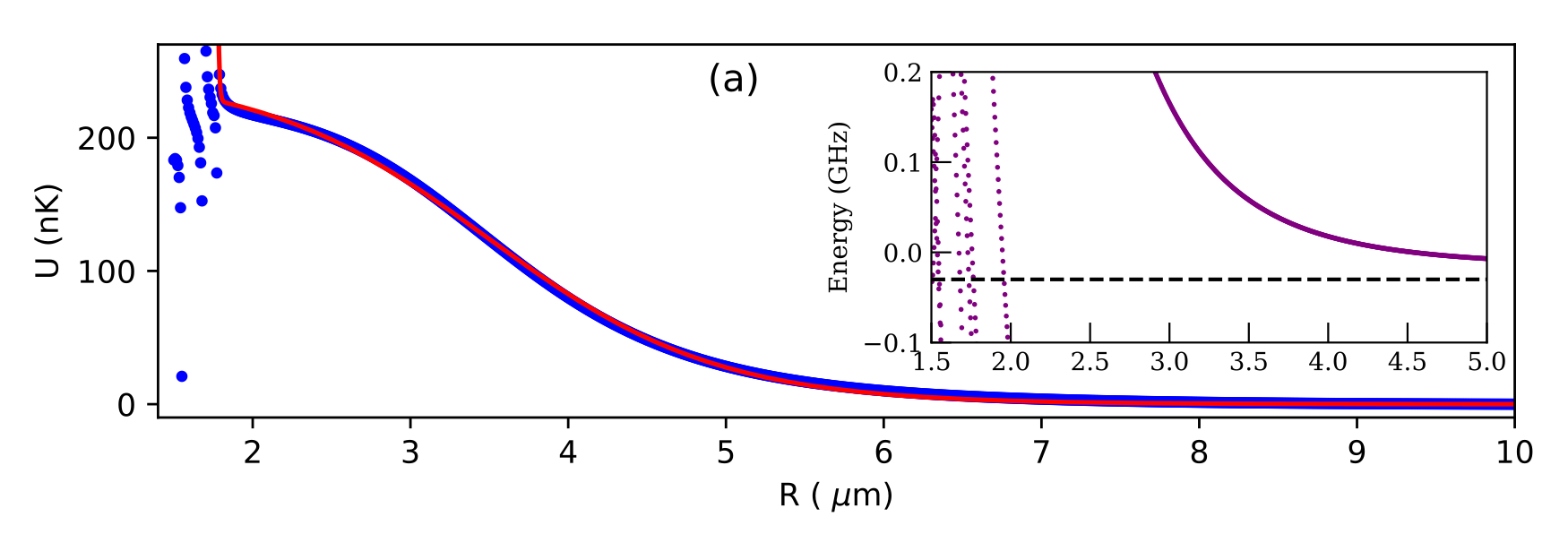}
    \includegraphics[width=0.75\textwidth]{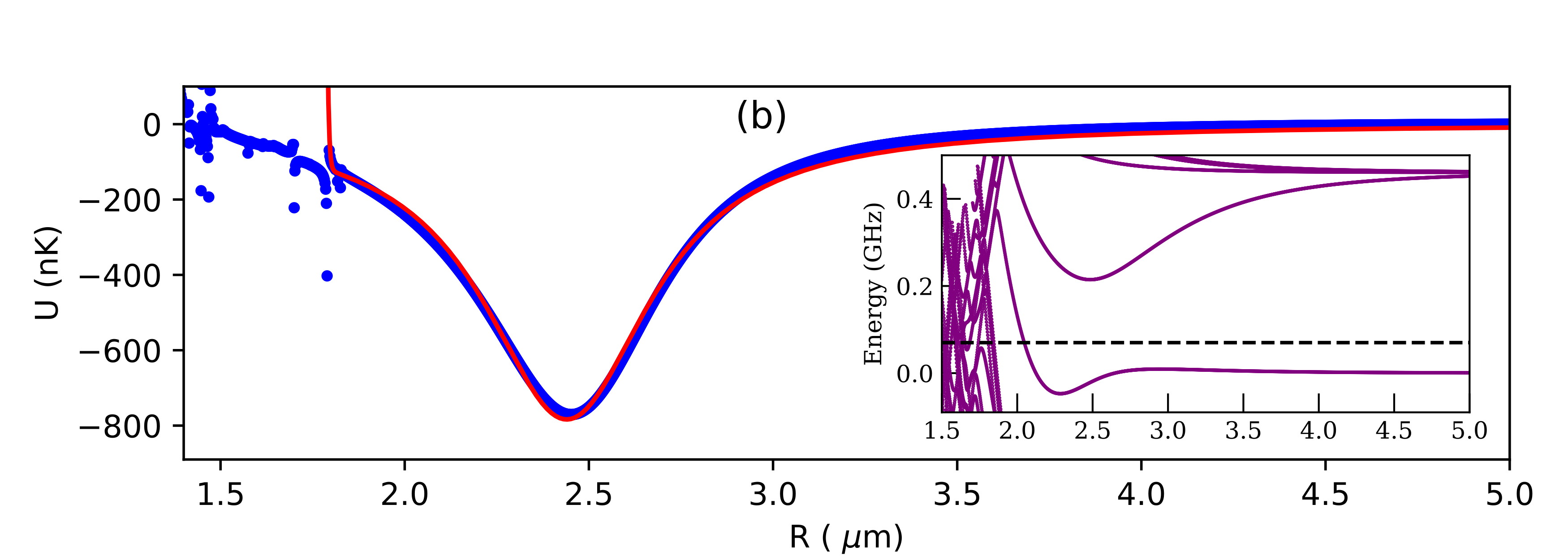}
    \caption{\textit{Top panel, (a):} The energy of the state dressed by $60S_{1/2}$ Rydberg level for $^{87}$Rb atoms as a function of interatomic distance. Here $\Omega_t/2\pi = 8$ MHz  and $\Delta_t/2\pi = 30$ MHz. The fit (red line) function is $248.9/\left(1+e^{1.382(x-3.494)}\right)+ e^{-145(x-1.811)}$. For the numerical diagonalization, 243 Rydberg states were used, and the Rydberg-Rydberg interaction included dipole-dipole, quadrupole-dipole and quadrupole-quadrupole terms. Inset shows how the interaction between Rydberg atoms changes as a function of interatomic distance. The dressed line shows where the dressed laser is resonant. \\
    \textit{Lower panel, (b):} The energy of the state dressed by $60P_{1/2}$ Rydberg level for $^{87}$Rb atoms as a function of interatomic distance. Here $\Omega_t/2\pi = 15$ MHz  and $\Delta_t/2\pi = -70$ MHz. The fit function is $783.6/\left(1+12.95(x-2.44)^2\right)+ e^{-263(x-1.813)}$. For the numerical diagonalization, 282 Rydberg states were used, and the Rydberg-Rydberg interaction included dipole-dipole, quadrupole-dipole and quadrupole-quadrupole terms. Inset shows how the interaction between Rydberg atoms changes as a function of interatomic distance. The dressed line shows where the dressed laser is resonant.}
    \label{fig_Ryd:eig_vals_60S}
\end{figure*}
In this section, we will discuss how we get the Rydberg dressed potentials we use in this work. To first order, the interaction between Rydberg dressed atoms can be easily calculated assuming a simple dipole-dipole interaction between atoms excited to Rydberg levels. In ref.~\cite{johnson_interactions_2010}, the authors calculated the potential between Rydberg-dressed atoms using such an approximation. They only consider a single Rydberg level, which results in an inter-atomic potential which scales as $1/R^3$ of resonant dipole-dipole interaction, where $R$ is the inter-atomic distance. In practice, as the Rydberg levels are closely spaced, multiple Rydberg levels have to be considered. In our analysis, we consider multiple Rydberg levels. In general, the Hamiltonian for two atoms interacting with a light field nearly resonant with a Rydberg level can be written as,
\begin{widetext}
\begin{equation} 
\begin{split}
H(t) =& \hbar\left[ \sum_{i}\omega_{i}(\ket{g,r_i}\bra{g,r_i}+\ket{r_i,g}\bra{r_i,g})  
     + \sum_{i,j}(\omega_{i}+\omega_{j})(\ket{r_j,r_i}\bra{r_j,r_i}+\ket{r_i,r_j}\bra{r_i,r_j}) \right.  \\
& + \left. \sum_{i}2\Omega_{i}e^{-i \omega_\text{L} t}(\ket{g,g}\bra{g,r_i}+\ket{g,g}\bra{r_i,g})+2\Omega^*_{i}e^{i \omega_\text{L} t}(\ket{g,r_i}\bra{g,g}+\ket{r_i,g}\bra{g,g})\right. \\
& + \left. \sum_{i,j}2\Omega_{i}e^{-i \omega_\text{L} t}(\ket{g,r_j}\bra{r_i,r_j}+\ket{r_j,g}\bra{r_j,r_i})+2\Omega^*_{i}e^{i \omega_\text{L} t}(\ket{r_i,r_j}\bra{g,r_i}+\ket{r_j,r_i}\bra{r_i,g})\right. \\
& + \left.  \sum_{i,j,k,l}D_{ijkl}(R)\ket{r_i,r_j}\bra{r_k,r_l}+D^*_{ijkl}(R)\ket{r_k,r_l}\bra{r_i,r_j}\right].
\end{split}
\end{equation}
\end{widetext}
Here the state, $\ket{g,r_i}$ denotes that the first atom is its ground state ($\ket{g}$) and the second atom is in the $i^\text{th}$ Rydberg state ($\ket{r_i}$). $\hbar\omega_{i}$ is the energy of the $i^\text{th}$ Rydberg state with respect to the ground state, $\omega_{L}/(2 \pi)$ is the frequency of the dressing laser, $\Omega_{i}/(2 \pi)$ is the Rabi frequency induced by the dressing laser for the transition $\ket{g}\rightarrow\ket{r_i}$, and $D_{ijkl}(R)$ is the interaction between rydberg level pairs $(r_i,r_j)$ and $(r_k,r_l)$ as a function of interatomic distance $R$.  

We can go to a rotating frame of reference using the following unitary transformation,
\begin{equation} 
\begin{split}
U(t) &= \ket{g,g}\bra{g,g} +  \sum_{i}e^{-i \omega_\text{L} t}(\ket{g,r_i}\bra{g,r_i}+\ket{r_i,g}\bra{r_i,g}) \\
&+ \sum_{ij}e^{-2i \omega_\text{L} t}(\ket{r_i,r_j}\bra{r_i,r_j}+\ket{r_j,r_i}\bra{r_j,r_i}).
\end{split}
\end{equation}
The Hamiltonian in this rotating frame becomes, 
\begin{widetext}
\begin{equation} 
\begin{split}
H_r =& U(t)H(t)U^{\dagger}(t) + i \hbar  U(t) \ dU^{\dagger}(t)/dt \\ 
=&\hbar\left[ \sum_{i}\Delta_{i}(\ket{g,r_i}\bra{g,r_i}+\ket{r_i,g}\bra{r_i,g})  
+ \sum_{i,j}(\Delta_{i}+\Delta_{j})(\ket{r_j,r_i}\bra{r_j,r_i}+\ket{r_i,r_j}\bra{r_i,r_j}) \right.  \\
& + \left. \sum_{i}2\Omega_{i}(\ket{g,g}\bra{g,r_i}+\ket{g,g}\bra{r_i,g})+2\Omega^*_{i}(\ket{g,r_i}\bra{g,g}+\ket{r_i,g}\bra{g,g})\right. \\
& + \left. \sum_{i,j}2\Omega_{i}(\ket{g,r_j}\bra{r_i,r_j}+\ket{r_j,g}\bra{r_j,r_i})+2\Omega^*_{i}(\ket{r_i,r_j}\bra{g,r_i}+\ket{r_j,r_i}\bra{r_i,g})\right. \\
& + \left.  \sum_{i,j,k,l}D_{ijkl}(R)\ket{r_i,r_j}\bra{r_k,r_l}+D^*_{ijkl}(R)\ket{r_k,r_l}\bra{r_i,r_j}\right],
\end{split}
\end{equation}
\end{widetext}
where $\Delta_{i} = \omega_i - \omega_\text{L}$ is the detuning of the laser from the transition formed by $\ket{g}$ and $\ket{r_i}$. 
We are interested in coupling the laser light to a single Rydberg state $r_\text{t}$, where $t$ denotes the target state. To do this we choose a laser wavelength such that $\Delta_i >> \Delta_r$ for all $i \ne t$. In this case we can neglect all levels $\ket{g,r_{i\ne t}}$ and $\ket{r_{i\ne t},g}$ as such states will not be excited. We keep the levels $\ket{r_i,r_j}$ as $D_{ijkl}(R)$ is of the same order of magnitude as $\Delta_{t}$ and that's where the $R$ dependence will come from. In such a scenario the Hamiltonian reduces to,
\begin{widetext}
\begin{equation} 
\begin{split}
H_r =&\hbar\left[ \Delta_{t}(\ket{g,r_t}\bra{g,r_t}+\ket{r_t,g}\bra{r_t,g})  
+ \sum_{i,j}(\Delta_{i}+\Delta_{j})(\ket{r_j,r_i}\bra{r_j,r_i}+\ket{r_i,r_j}\bra{r_i,r_j}) \right.  \\
& + \left. 2\Omega_{i}(\ket{g,g}\bra{g,r_t}+\ket{g,g}\bra{r_t,g})+2\Omega^*_{i}(\ket{g,r_t}\bra{g,g}+\ket{r_t,g}\bra{g,g})\right. \\
& + \left. \sum_{i}2\Omega_{i}(\ket{g,r_t}\bra{r_i,r_t}+\ket{r_t,g}\bra{r_t,r_i})+2\Omega^*_{i}(\ket{r_i,r_t}\bra{g,r_t}+\ket{r_t,r_i}\bra{r_t,g})\right. \\
& + \left.  \sum_{i,j,k,l}D_{ijkl}(R)\ket{r_i,r_j}\bra{r_k,r_l}+D^*_{ijkl}(R)\ket{r_k,r_l}\bra{r_i,r_j}\right].
\end{split}
\end{equation}
\end{widetext}

We numerically diagonalize the above Hamiltonian to get the dressed eigenstate and extract the eigenvalue closest to the state $\ket{g,g}$. For an atom dressed primarily with the $60S_{1/2}$ Rydberg state of $^{87}$Rb atom, the lowest eigenvalue as a function of the inter-atomic distance is shown in Fig.~\ref{fig_Ryd:eig_vals_60S}~(a). Similarly, if the dressing laser is near the $60P_{1/2}$ Rydberg level, we get a dressed potential as shown in Fig.~\ref{fig_Ryd:eig_vals_60S}~(b).
For the numerical diagonalization, we modified the Python library named ARC~\cite{sibalic_arc_2017} to include the dressing field.

\bibliography{bibliography.bib}

\end{document}